\documentclass[prd,showpacs,amsmath,amssymb]{revtex4}
\usepackage{dcolumn}
\usepackage{bm}
\usepackage{epsfig,graphicx}
\newcommand{\be}{\begin{equation}}
\newcommand{\ee}{\end{equation}}
\newcommand{\bea}{\begin{eqnarray}}
\newcommand{\eea}{\end{eqnarray}}
\newcommand{\bean}{\begin{eqnarray*}}
\newcommand{\eean}{\end{eqnarray*}}

\newcommand{\gapproxeq}{\lower
.7ex\hbox{$\;\stackrel{\textstyle >}{\sim}\;$}}
\newcommand{\lapproxeq}{\lower
.7ex\hbox{$\;\stackrel{\textstyle <}{\sim}\;$}}

\begin{document}

\bibliographystyle{unsrt}

\title{\bf Further understanding of the non-$D\bar D$ decays of $\psi(3770)$}

\author{Gang Li$^{1}$~\footnote{Email:gli@mail.qfnu.edu.cn},
Xiao-Hai Liu$^{2,3}$~\footnote{Email:liuxiaohai@pku.edu.cn}, Qian
Wang$^{3,4}$~\footnote{Email:q.wang@fz-juelich.de}, and Qiang
Zhao$^{3,5}$~\footnote{Email:zhaoq@ihep.ac.cn}}

\affiliation{1) Department of Physics, Qufu Normal University, Qufu
273165, P.R. China}

\affiliation{2) Department of Physics and State Key Laboratory of Nuclear Physics and Technology, Peking University, Beijing 100871, China}

\affiliation{3) Institute of High Energy Physics, Chinese Academy of
Sciences, Beijing 100049, P.R. China}

\affiliation{4) Institut f\"{u}r Kernphysik and J\"ulich Center for
Hadron Physics, Forschungszentrum J\"{u}lich, D--52425 J\"{u}lich,
Germany}

\affiliation{5) Theoretical Physics Center for Science Facilities,
Chinese Academy of Sciences, Beijing 100049, P.R. China}

\date{\today}

\begin{abstract}
We provide details of the study of $\psi(3770)$ non-$D\bar D$ decays
into $VP$, where $V$ and $P$ denote light vector meson and
pseudoscalar meson, respectively. We find that the electromagnetic
(EM) interaction plays little role in these processes, while the
strong interaction dominates. The strong interaction can be
separated into two parts, i.e. the short-distance part probing the
wave function at origin and the long-distance part reflecting the
soft gluon exchanged dynamics. The long-distance part is thus
described by the intermediate charmed meson loops. We show that the
transition of $\psi(3770)\to VP$ can be related to $\psi(3686)\to
VP$ such that the parameters in our model can be constrained by
comparing the different parts in $\psi(3770)\to VP$ to those in
$\psi(3686)\to VP$. Our quantitative results confirm the findings of
[Zhang {\it et al.}, Phys. Rev. Lett. 102, 172001 (2009)] that the
OZI-rule-evading long-distance strong interaction via the IML plays
an important role in $\psi(3770)$ decays, and could be a key towards
a full understanding of the mysterious $\psi(3770)$ non-$D\bar{D}$
decay mechanism.
\end{abstract}

\maketitle

\section{Introduction}
\label{Introduction}

The $\psi(3770)$ is the lowest mass charmonium resonance above the
open charm pair $D\bar D$ production threshold. Traditional theories
expect that it decays almost entirely into pure $D\bar D$ pairs
without the so-called Okubo-Zweig-Iizuka (OZI) rule~\cite{OZI}
suppression. An interesting and nontrivial question here is whether
the $\psi(3770)$ decay is totally saturated by $D\bar D$ , or
whether there exist significant non-$D\bar D$ decay channels. CLEO
measured the exclusive cross section $\sigma(e^+ e^-\to
 D^0\bar {D}^0)=(3.66\pm0.03\pm0.06) $ nb, $\sigma(e^+ e^-\to
 D^+D^-)=(2.91\pm0.03\pm0.05) $ nb at the center of mass energy $3774\pm 1 \mathrm{MeV}$~\cite{Dobbs:2007zt,He:2005bs} and the
inclusive cross section $\sigma(e^+ e^-\to \psi(3770)\to
\textrm{non-}D\bar D)= -0.01\pm 0.08^{+0.41}_{-0.30}$
nb~\cite{Besson:2005hm}. These results lead to
$\textrm{BR}(\psi(3770)\to D\bar{D})=(103.0\pm 1.4
^{+5.1}_{-6.8})\%$, of which the lower bound suggests the maximum
non-$D\bar{D}$ branching ratio is about 6.8\%.

BES earlier reported two results based on different analysis
methods: $\textrm{BR}(\psi(3770)\to \textrm{non-}D\bar D) = (14.5\pm
1.7\pm 5.8)\%$ \cite{Ablikim:2006aj} and $\textrm{BR}(\psi(3770)\to
\textrm{non-} D\bar D) = (16.4\pm 7.3\pm 4.2)\%$ \cite{He:2005bs}.
With the first direct measurement on the non-$D\bar D$ decay, BES
gives $\sigma(e^{+}e^{-}\rightarrow\psi(3770)\rightarrow
\textrm{non-}D\bar D)=(0.95\pm0.35\pm0.29)$ nb and
$\textrm{BR}(\psi(3770)\rightarrow\textrm{non-}D\bar{D})=(13.4\pm5.0\pm3.6)\%$~\cite{Ablikim:2007zz}.
Evidently, the two collaborations give very different results of the
non-$D\bar D$ decay of $\psi(3770)$. Meanwhile, various exclusive
decay channels have been investigated by
BES~\cite{Ablikim:2004bc,Ablikim:2005cd,Ablikim:2007ss,:2007wg,:2007ht}
and CLEO~\cite{Huang:2005fx,Adams:2005ks,CroninHennessy:2006su} in
order to search for further non-$D\bar D$ decay modes of
$\psi(3770)$.

In Ref.~\cite{Nakamura:2010zzi}, several non-$D\bar{D}$ hadronic
decay branching ratios are listed, i.e. $\psi(3770)\to J/\psi
\pi^+\pi^-$, $J/\psi \pi^0\pi^0$, $J/\psi\eta$ and $\phi\eta$, while
tens of other channels have only experimental upper limits due to
the poor statistics.  These results are within the range of
theoretical predictions based on the QCD multipole expansion for
hadronic transitions~\cite{Kuang:1989ub}. With the total width of
$27.3\pm 1.0$~MeV for $\psi(3770)$~\cite{Nakamura:2010zzi}, the
width of all hadronic transitions is about $100-150$ KeV. Another
kind of non-$D\bar D$ decays of $\psi(3770)$ are the E1 transitions
$\psi(3770)\rightarrow \gamma \chi_{c0}$ and $\gamma \chi_{c1}$,
their branching ratios are measured to be $(7.3\pm 0.9)\times
10^{-3}$ and $(2.9\pm 0.6) \times 10^{-3}$,
respectively~\cite{Nakamura:2010zzi}, while only an upper limit is
given to $\psi(3770) \to \gamma\chi_{c2}$. The sum of those
channels, however, is far from clarifying the mysterious situation
of the $\psi(3770)$ non-$D\bar{D}$ decays. It hence stimulates
intensive experimental and theoretical
efforts~\cite{Kuang:1989ub,Ding:1991vu,Rosner:2001nm,Rosner:2004wy,Eichten:2007qx,Voloshin:2005sd,Achasov:1990gt,Achasov:1991qp,Achasov:1994vh,Achasov:2005qb,He:2008xb}
on understanding the nature of $\psi(3770)$ and its strong and
radiative transition mechanisms.

Theoretically, Kuang and Yan~\cite{Kuang:1989ub} calculated the
transition $\psi(3770)\to J/\psi \pi\pi$ with the QCD multipole
expansion, which indicated small exclusive non-$D\bar{D}$ decays
width at leading order.  Recently, He, Fan and Chao~\cite{He:2008xb}
calculate the light hadron decays of $\psi(3770)$ based on
nonrelativistic QCD factorization at next-to-leading order (NLO) in
$\alpha_s$ and leading order in~$v^2$, which gives $\Gamma
[\psi(3770) \to \textrm{light hadrons}]= 467_{+338}^{-187}$ keV.
Although the NLO contributions are found important, it has also been
pointed out in Ref.~\cite{He:2008xb} that any significant
non-$D\bar{D}$ branching ratios up to a few percent would be likely
due to non-perturbative mechanisms instead of QCD higher order
contributions. Since the mass of $\psi(3770)$ is close to the
$D\bar{D}$ threshold, a natural mechanism involving non-perturbative
mechanisms is the $D\bar{D}$ final state interaction. Namely, the
rescattering of the $D\bar{D}$ as intermediate states can contribute
to final state non-$D\bar{D}$ decay channels via charmed meson
loops. Such a mechanism has been quantified in
Ref.~\cite{Zhang:2009kr} for $\psi(3770)\to VP$, where the
intermediate meson loops (IML) are extended to include
$D\bar{D}^*+c.c.$ contributions. Since the mass threshold of
$D\bar{D}^*+c.c.$ is higher than the mass of $\psi(3770)$, it should
be noted that only the intermediate $D\bar{D}$ can contribute to the
absorptive part of the non-$D\bar{D}$ transition amplitude, while
other IML will contribute to the dispersive part. Taking into
account that the $\psi(3770)$ is coupled to the $D\bar{D}$ pair by a
$P$ wave, we also note that model-dependence of the IML calculations
will become inevitable when evaluate the dispersive amplitudes. In
Ref.~\cite{Liu:2009dr} the same mechanism is also investigated for
$\psi(3770) \to J/\psi \eta$, $\rho\pi$, $\omega\eta$ and $K^*K$,
which is consistent with the result of Ref.~\cite{Zhang:2009kr}.

The impact of the IML as an important non-perturbative mechanism has
been extensively studied in charmonium decays during the past few
years. It indicates  that since the charm quark is not heavy enough
the decay of charmonium states would be sensitive to the
interferences between the perturbative and non-perturbative
mechanisms in the charmonium energy region. With more and more data
from Belle, BaBar, CLEO and BES-III, it is broadly recognized that
the intermediate hadron loops can be closely related to a lot of
non-perturbative phenomena observed in experiment, e.g. apparent
OZI-rule
violations~\cite{Zhang:2009kr,Liu:2009vv,Liu:2010um,Wang:2010iq,Guo:2010zk,Guo:2010ak,Zhao:2005ip,Wu:2007jh,Wang:2012mf,Cheng:2004ru,Anisovich:1995zu,Li:2007au,Li:1996yn,Zhao:2006dv,Zhao:2006cx},
helicity selection rule
violations~\cite{Liu:2009vv,Liu:2010um,Wang:2010iq}, and isospin
symmetry breakings in charmonium
decays~\cite{Guo:2012tj,Wang:2011yh}. In this work, we present a
more detailed analysis of the $\psi(3770)$ non-$D\bar{D}$ decays
into a pair of vector and pseudoscalar mesons. In particular, we try
to constrain the model parameters such that a quantitative
evaluation of the decay branching ratios of $\psi(3770)\to VP$ can
be achieved.


This paper is organized as below. In Sec. II we present the
framework of our calculation. Section III is devoted to numerical
results and discussions. A brief summary is given in the last
section.

\section{The framework}
\label{method}

The IML contributions in $\psi(3770)$ can be related to the similar
processes of $\psi(3686)\to VP$ as discussed in
Ref.~\cite{Wang:2012mf}. This is understandable since both
$\psi(3770)$ and $\psi(3686)$ are located in the vicinity of the
$D\bar{D}$ threshold. Thus, their decays would experience similar
effects from the $D\bar{D}$ threshold. Such a conjecture allows us
to correlate these two states together such that some of the
parameters can be fixed for both cases. Meanwhile, we stress some
differences between these two decays, i.e. $\psi(3770)$ and
$\psi(3686)\to VP$. For $\psi(3770)$ the intermediate $D\bar{D}$
meson loop can contribute to both the absorptive and dispersive part
of the transition amplitude, while for $\psi(3686)$ the IML will
only contribute to the dispersive part since $\psi(3686)$ is below
the $D\bar{D}$ threshold.

Benefiting from the unique antisymmetric tensor structure for the
$VVP$ coupling, we can decompose the transition amplitude into three
parts, i.e. the EM interaction, short-distance strong interaction
and long-distance strong interaction.

\subsection{The electromagnetism contribution}
\begin{figure}
\begin{center}
\includegraphics[scale=0.7]{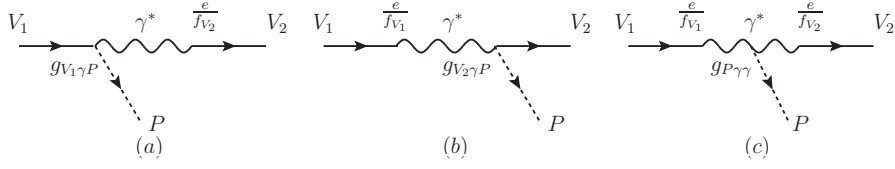}
\vspace{-16cm} \caption{The tree-level Feynman diagrams of EM
transitions in $\psi(3770)\to VP$.}
 \label{EM}
\end{center}
\end{figure}
As shown in Fig.~\ref{EM}, the tree level amplitudes of the EM part
can be described by the vector meson dominance (VMD),
\begin{eqnarray}
\nonumber
\mathcal{M}_{EM}&=&\mathcal{M}_a+\mathcal{M}_b+\mathcal{M}_c\\
&=&\left(\frac{e}{f_{V_2}}\frac{g_{V_1\gamma
P}}{M_{V_1}}\mathcal{F}_a+\frac{e}{f_{V_1}}\frac{g_{V_2\gamma
P}}{M_{V_2}}\mathcal{F}_b+\frac{e^2}{f_{V_1}f_{V_2}}\frac{g_{P\gamma\gamma
}}{M_{P}}\mathcal{F}_c\right)\epsilon_{\mu\nu\alpha\beta}p^\mu\epsilon(p)^\nu
k^\alpha\epsilon(k)^\beta,
\end{eqnarray}
where $p$($k$) is the four momentum of the initial vector charmonium
(final light vector), and $\epsilon(p)$ ($\epsilon(k)$) is its
corresponding polarization vector. The effective couplings in the
amplitudes can be found in Table. I-III of Ref.~\cite{Wang:2012mf}.
The effective coupling $e/f_{\psi(3770)}=5.4\times
10^{-3}$~\cite{Wang:2010iq} is used here. Meanwhile, the upper
limits of the branching ratios $\mathrm{BR}(\psi(3770)\to
\gamma\pi^0)< 2\times 10^{-4}$, $\mathrm{BR}(\psi(3770)\to
\gamma\eta)< 1.5\times 10^{-4}$ and $\mathrm{BR}(\psi(3770)\to
\gamma\eta^\prime)< 1.8\times 10^{-4}$ are used to extract the
effective couplings $g_{\gamma\psi(3770)\pi^0}=4.38\times 10^{-4}$,
$g_{\gamma\psi(3770)\eta}=3.49\times 10^{-4}$ and
$g_{\gamma\psi(3770)\eta^\prime}=4.8\times 10^{-4}$. Considering the
off-shell photon, a monopole form factor
$\mathcal{F}(q^2)=\Lambda_{EM}^2/(\Lambda_{EM}^2-q^2)$ is included.
Since the mass difference between $\psi(3770)$ and $\psi(3686)$ is
rather small,  we adopt the same cut off parameter
$\Lambda_{EM}=0.542 \ \mathrm{GeV}$ as in $\psi(3686)$
decays~\cite{Wang:2012mf} to evaluate the EM contribution.  Namely,
the EM contribution can be constrained by taking into account the
$\psi(3686)$ transitions.

\subsection{The short-distance strong interaction}
\begin{figure}
\begin{center}
\includegraphics[scale=0.7]{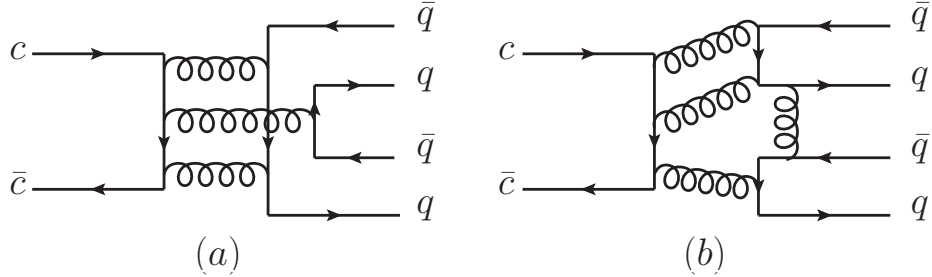}
\vspace{-15cm} \caption{Schematic diagrams for the short-distance
strong transitions in $\psi(3770)\to VP$. Diagram (a) illustrates
the SOZI process, while (b) is for the DOZI one. In both cases,  $c$
and $\bar{c}$ annihilate at the origin of the wavefunction. }
 \label{short}
\end{center}
\end{figure}

As shown in Fig.~\ref{short}, the short-distance strong interaction
is $c\bar{c}$ annihilation into three hard gluons. We use the simple
parameterized scheme~\cite{Wang:2012mf} to describe this part. The
amplitudes of the $VP$ channels can be written as
\begin{eqnarray}
\mathcal{M}_S(\rho^0\pi^0)&=&
\mathcal{M}_S(\rho^+\pi^-)=\mathcal{M}_S(\rho^-\pi^+)=g{\cal
{F}}({\bf P}) \ , \nonumber \\
\mathcal{M}_S(K^{*+}K^-)&=&
\mathcal{M}_S(K^{*-}K^+)=\mathcal{M}_S(K^{*0}\bar{K^0})=\mathcal{M}_S(\bar{K^{*0}}K^0)
=g\xi {\cal {F}}({\bf P})  \ , \nonumber\\
\mathcal{M}_S(\omega\eta)&=&X_\eta g(1+2r){\cal
{F}}({\bf P})+Y_\eta\sqrt{2}\xi rg{\cal {F}}({\bf P})  \ , \nonumber\\
\mathcal{M}_S(\omega\eta^\prime)&=&X_{\eta^\prime} g(1+2r){\cal
{F}}({\bf P})+Y_{\eta^\prime}\sqrt{2}\xi rg{\cal {F}}({\bf P})  \ , \nonumber\\
\mathcal{M}_S(\phi\eta)&=& X_{\eta}\sqrt{2}\xi rg{\cal {F}}({\bf
P})+Y_\eta g(1+r)\xi^2 {\cal
{F}}({\bf P})  \ , \nonumber\\
\mathcal{M}_S(\phi\eta^\prime)&=& X_{\eta^\prime}\sqrt{2}\xi rg{\cal
{F}}({\bf P})+Y_{\eta^\prime} g(1+r)\xi^2 {\cal {F}}({\bf P}) ,
\end{eqnarray}
where $g$, $\xi$ and $r$ are the transition strength of the singly
disconnected OZI (SOZI) process, SU(3) breaking parameter and doubly
disconnected OZI (DOZI) process parameter~\cite{Wang:2012mf}. ${\cal
{F}}({\bf P}) \equiv |{\bf P}|^{l}\exp ({-{\bf P}^2/{16\beta^2}})$
(with $\beta=0.5 \ \mathrm{GeV}$) is the exponential form factor
which is obtained from the chiral constitute quark model reflecting
the size effect of the initial and final particles.

The $\eta$-$\eta^\prime$ mixing is considered in a standard way,
\bea
\eta &=& \cos\alpha_P |n\bar n\rangle  - \sin\alpha_P |s\bar
s\rangle,
\nonumber\\
\eta^\prime &=& \sin\alpha_P |n\bar n\rangle + \cos\alpha_P |s \bar
s\rangle ,
\eea
where $|n\bar{n}\rangle \equiv |u\bar{u}
 +d\bar{d}\rangle/\sqrt{2}$, and the mixing angle $\alpha_P=\theta_P +
\arctan(\sqrt{2})$ with $\theta_P\simeq -24.6^\circ$ or $\sim
-11.5^\circ$ for linear or quadratic mass relations,
respectively~\cite{Nakamura:2010zzi}.  We adopt $\theta_P =
-22^\circ$ ~\cite{Wang:2012mf}. Here,
$X_\eta=Y_{\eta^\prime}=\cos\alpha_P$ and
$-Y_\eta=X_{\eta^\prime}=\sin\alpha_P$.

Again, since the mass difference between $\psi(3770)$ and
$\psi(3686)$  is not large, the SU(3) breaking effect and DOZI
parameter can be taken the same as those determined in the
$\psi(3686)$ decays, i.e. $\xi=0.92$, $r=-0.097$. Meanwhile, the
wave function at origin can be extracted from the lepton pair decay
width. The branching ratios $BR(\psi(3686)\to e^+e^-)=7.72\times
10^{-3}$ and $BR(\psi(3770)\to e^+e^-)=9.7\times 10^{-6}$ gives the
ratio of the wave function at origin
$\frac{\psi(3770)(0)}{\psi(3686)(0)}=0.35$, implying
$g_{\psi(3770)}\simeq 0.35\times g_{\psi(3686)}=0.35\times
1.25\times10^{-2}=4.38\times 10^{-3}$. Under this circumstance, the
short-distance strong contribution is also determined.

\subsection{The long-distance strong interaction }

\begin{figure}[tb]
\begin{center}\vglue-0mm
\includegraphics[width=0.8\textwidth]{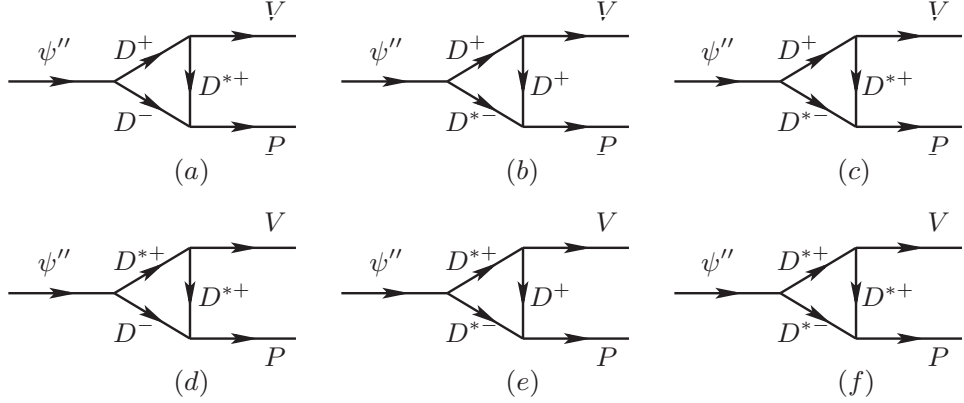}
\vglue-0mm\caption{Diagrams of charged intermediate meson loops
contributing to $\psi(3770)\to VP$, where $V$ and $P$ denote light
vector and pseudoscalar mesons, respectively. Similar diagrams with
the neutral and strange charmed meson loops are
implicated.}\label{long}
\end{center}
\end{figure}

Similar to Ref.~\cite{Zhang:2009kr}, the long-distance contributions
are considered by the IML as illustrated in Fig.~\ref{long}. As
studied in Ref.~\cite{Zhang:2009kr}, the $s$-channel process via
$2S-1D$ mixing does not contribute significantly to the amplitude.
Therefore, only the $t$-channel transitions are studied in the
evaluation. In addition, an improvement here is that the form factor
is introduced in line with the study of the $\psi(3686)$
decays~\cite{Wang:2012mf} such that a correlation between these two
states can be established and provide more stringent constraints on
the underlying mechanisms. Since the non-local effects cannot be
well controlled in the ELA framework, such a correlation would be
crucially useful for quantifying the IML as a possible
non-perturbative mechanism in charmonium decays.

The couplings between an $S$-wave charmonium and charmed mesons are
given by the effective Lagrangian based on heavy quark
symmetry~\cite{Colangelo:2003sa,Casalbuoni:1996pg},
\begin{equation}
\mathcal{L}_2=i g_2 \mathrm{Tr}[R_{c\bar{c}} \bar{H}_{2i}\gamma^\mu
{\stackrel{\leftrightarrow}{\partial}}_\mu \bar{H}_{1i}] + H.c.,
\end{equation}
where the $S$-wave vector and pseudoscalar charmonium states are
expressed as
\begin{equation}
R_{c\bar{c}}=\left( \frac{1+ \rlap{/}{v} }{2} \right)\left(\psi^\mu
\gamma_\mu -\eta_c \gamma_5 \right )\left( \frac{1- \rlap{/}{v} }{2}
\right).
\end{equation}
The charmed and anti-charmed meson triplet are
\begin{eqnarray}
H_{1i}&=&\left( \frac{1+ \rlap{/}{v} }{2} \right)[
\mathcal{D}_i^{*\mu}
\gamma_\mu -\mathcal{D}_i\gamma_5], \\
H_{2i}&=& [\bar{\mathcal{D}}_i^{*\mu} \gamma_\mu
-\bar{\mathcal{D}}_i\gamma_5]\left( \frac{1- \rlap{/}{v} }{2}
\right),
\end{eqnarray}
where $\mathcal{D}$ and $\mathcal{D}^*$ are pseudoscalar
($(D^{0},D^{+},D_s^{+})$) and vector charmed mesons
($(D^{*0},D^{*+},D_s^{*+})$), respectively.

Explicitly, the Lagrangian for $\psi(3770)$ couplings to $D$ and
$D^*$ becomes
\begin{eqnarray}\label{LS}
\mathcal{L}_\psi &=& ig_{\psi \mathcal{D}^* \mathcal{D}^*} (-\psi^\mu
\mathcal{D}^{*\nu}\overleftrightarrow{\partial}_\mu
\mathcal{D}_\nu^{*\dagger}+ \psi^\mu
\mathcal{D}^{*\nu}\partial_\nu\mathcal{D}^{*\dagger}_{\mu} -
\psi_\mu\partial_\nu \mathcal{D}^{*\mu} \mathcal{D}^{*\nu\dagger})\nonumber\\
&&+ ig_{\psi \mathcal{D}\mathcal{D}}\psi_\mu(\partial^\mu
\mathcal{D}\mathcal{D}^{\dagger}-\mathcal{D}\partial^\mu
\mathcal{D}^{\dagger}) + g_{\psi
\mathcal{D}^*\mathcal{D}}\varepsilon^{\mu\nu\alpha\beta}\partial_\mu
\psi_\nu(\mathcal{D}^*_\alpha\overleftrightarrow{\partial}_\beta
\mathcal{D}^{\dagger}
- \mathcal{D}\overleftrightarrow{\partial}_\beta\mathcal{D}^{*\dagger}_\alpha) ,
\end{eqnarray}
with
\begin{eqnarray}
{g_{\psi DD}}={g_{\psi D^*D^*}}, \, \,\,\, \, g_{\psi D^*D} = \frac
{g_{\psi DD}} {\sqrt {m_{D^*} m_D}}.
\end{eqnarray}

The coupling $g_{\psi(3770) D \bar D}$ is extracted with the help of
the experimental data~\cite{Nakamura:2010zzi},
 \be \Gamma_{\psi(3770)\to D \bar D} =
\frac{g_{\psi(3770) D \bar D}^2 |\vec{p}\,|^3}{6\pi
M_{\psi(3770)}^2},
 \ee
 where $|\vec{p}\,|$ is the three momentum of final D-meson. The branching ratios
for $\psi(3770) \to D^+ D^-$ and $D^0 \bar D^0$ are slightly
different, which gives $g_{\psi(3770) D^+ D^-} =13.55$ and $
g_{\psi(3770) D^0 \bar D^0}=12.69$. The averaged value 13.12 is used
in our calculation.

The Lagrangians relevant to the light vector and pseudoscalar mesons
are,
 \begin{eqnarray}
 {\cal L} &=& - ig_{D^*DP}(D^i\partial^\mu P_{ij}
 D_\mu^{*j\dagger}-D_\mu^{*i}\partial^\mu P_{ij}D^{j\dagger})
 +{1\over 2}g_{D^*D^*P}
 \varepsilon_{\mu\nu\alpha\beta}\,D_i^{*\mu}\partial^\nu P^{ij}
 {\overleftrightarrow \partial}{}^{\!\alpha} D^{*\beta\dagger}_j \nonumber \\
 &-& ig_{DDV} D_i^\dagger {\overleftrightarrow \partial}_{\!\mu} D^j(V^\mu)^i_j
 -2f_{D^*DV} \epsilon_{\mu\nu\alpha\beta}
 (\partial^\mu V^\nu)^i_j
 (D_i^\dagger{\overleftrightarrow \partial}{}^{\!\alpha} D^{*\beta j}-D_i^{*\beta\dagger}{\overleftrightarrow \partial}{}{\!^\alpha} D^j)
 \nonumber\\
 &+& ig_{D^*D^*V} D^{*\nu\dagger}_i {\overleftrightarrow \partial}_{\!\mu} D^{*j}_\nu(V^\mu)^i_j
 +4if_{D^*D^*V} D^{*\dagger}_{i\mu}(\partial^\mu V^\nu-\partial^\nu
 V^\mu)^i_j D^{*j}_\nu,
 \label{eq:LDDV}
 \end{eqnarray}
with the convention $\varepsilon_{0123}=1$, where $P$ and $V_\mu$
are $3\times 3$ matrices for the nonet pseudoscalar and vector
mesons, respectively,
 \begin{eqnarray}  P &= &\left(
\begin{array} {ccc}
 \frac {\pi^0} {\sqrt {2}} +{ \eta\cos\alpha_P+\eta^\prime\sin\alpha_P\over\sqrt{2}} & \pi^+ & K^+ \\
\pi^- & -\frac {\pi^0} {\sqrt {2}} +{ \eta\cos\alpha_P +  \eta^\prime\sin\alpha_P\over\sqrt{2}} & K^0 \\
K^-& {\bar K}^0 & - \eta\sin\alpha_P + \eta^\prime{\cos\alpha_P} \\
\end{array}\right), \\
V &= &\left(\begin{array}{ccc}\frac{\rho^0} {\sqrt {2}}+\frac {\omega} {\sqrt {6}}&\rho^+ & K^{*+} \\
\rho^- & -\frac {\rho^0} {\sqrt {2}} + \frac {\omega} {\sqrt {2}} & K^{*0} \\
K^{*-}& {\bar K}^{*0} & \phi \\
\end{array}\right).
\end{eqnarray}

In the chiral and heavy quark limits, the charmed meson couplings to
light meson have the following
relationship~\cite{Casalbuoni:1996pg},
\begin{eqnarray} g_{D^\ast
D^\ast \pi} = \frac{g_{D^\ast D \pi}}{\sqrt{m_D m_{D^\ast}}} =
\frac{2}{ f_\pi } g ,\quad g_{DDV}=g_{D^*D^*V}=\frac{\beta
g_V}{\sqrt2}, \quad
f_{D^*DV}&=&\frac{f_{D^*D^*V}}{m_{D^*}}=\frac{\lambda g_V}{\sqrt2},
\end{eqnarray}
where $f_\pi$ = 132 MeV is the pion decay constant, the parameters
$g_V$ is given by $g_V = {m_\rho / f_\pi}$~\cite{Casalbuoni:1996pg}.
We take $\lambda = 0.56\, \text{GeV}^{-1} $ and $g =
0.59$~\cite{Isola:2003fh} in the calculation.

The loop transition amplitudes in  Fig.~\ref{long} can be expressed
in a general form in the ELA as follows:
 \begin{eqnarray}
 M_{fi}=\int \frac {d^4 q_2} {(2\pi)^4} \sum_{D^\ast \ \mbox{pol.}}
 \frac {V_1V_2V_3} {a_1 a_2 a_3}\prod_i{\cal F}_i(m_i,q_i^2)
 \end{eqnarray}
 where $V_i \ (i=1,2,3)$ are the vertex functions;
 $a_i = q_i^2-m_i^2 \ (i=1,2,3)$ are the denominators of the intermediate meson propagators.
We adopt the tri-monopole form factor~\cite{Wang:2012mf},
$\prod_i{\cal F}_i(m_i,q_i^2)$, which is a product of off-shell
monopole form factors for each internal mesons, i.e.
\begin{equation}\label{ELA-form-factor}
\prod_i{\cal F}_i(m_i,q_i^2)\equiv {\cal F}_1(m_1,q_1^2){\cal
F}_2(m_2,q_2^2){\cal F}_3(m_3,q_3^2) \ ,
\end{equation}
where
\begin{equation}{\cal F}_i(m_{i}, q_i^2) \equiv \frac
{\Lambda_i^2-m_{i}^2} {\Lambda_i^2-q_i^2},
\end{equation}
with  $\Lambda_i\equiv m_i+\alpha\Lambda_{QCD}$ and the QCD energy
scale $\Lambda_{QCD} = 220$ MeV. This form factor will take into
account the non-local effects of the vertex functions and kill the
loop divergence in the integrals.

In principle, one should include all the possible
intermediate-meson-exchange loops in the calculation. In reality,
the breakdown of the local quark-hadron duality allows us to pick up
the leading contributions as a reasonable approximation
\cite{Lipkin:1986bi,Lipkin:1986av}. Also, intermediate states
involving flavor changes turn out to be strongly suppressed. One
reason is because of the large virtualities involved. The other is
because of the OZI rule suppressions. So we will only consider the
charmed meson loops here. The decay amplitudes for $\psi(3770)\to V
P$ corresponding to the diagrams in Fig.~\ref{long} read as
\begin{eqnarray}
{\cal M}_a&=&(i^3)\int \frac{d^4q_2} {(2\pi)^4}[g_{\psi DD}\varepsilon_\psi^\mu (q_{1\mu}-q_{3\mu})]
[-4if_{D^*DV} \varepsilon_{\rho\sigma\xi\tau}p_V^\rho\varepsilon_V^{*\sigma} q_1^\xi]
[2ig_{D^*DP}p_{P\phi} ] \nonumber \\
&& \times \frac {i} {q_1^2-m_1^2} \frac {i(-g^{\tau\phi}+q_2^\tau q_2^\phi/m_2^2)} {q_2^2-m_2^2}
\frac {i} {q_3^2-m_3^2}\prod_i {\cal F}_i(m_i, q_i^2) \ ,\\
{\cal M}_b&=&(i^3)\int \frac{d^4q_2} {(2\pi)^4}[g_{\psi
D^*D}\varepsilon_{\alpha\beta\mu\nu}p_\psi^\alpha\varepsilon_\psi^\beta
(q_3^\nu-q_1^\nu)] [-ig_{DDV}\varepsilon_V^{*\rho}(q_{1\rho}+q_{2\rho})] [ig_{D^*DP}p_{P\phi}]\nonumber\\
&& \times  \frac {i} {q_1^2-m_1^2} \frac {i} {q_2^2-m_2^2}
\frac {i(-g^{\mu\phi}+q_3^\mu q_3^\phi/m_3^2)} {q_3^2-m_3^2}\prod_i {\cal F}_i(m_i, q_i^2)  \ ,\\
{\cal M}_c&=&(i^3)\int \frac{d^4q_2} {(2\pi)^4}[g_{\psi
D^*D}\varepsilon_{\alpha\beta\mu\nu}p_\psi^\alpha\varepsilon_\psi^\beta
(q_3^\nu-q_1^\nu)] [-4if_{D^*DV} \varepsilon_{\rho\sigma\xi\tau}p_V^\rho\varepsilon_V^{*\sigma} q_1^\xi]
[-i g_{D^*D^*P} \varepsilon_{\phi\lambda\theta\kappa} p_P^\lambda q_2^\theta ]\nonumber\\
&& \times  \frac {i} {q_1^2-m_1^2} \frac {i(-g^{\tau\kappa}+q_2^\tau q_2^\kappa/m_2^2)} {q_2^2-m_2^2}
\frac {i(-g^{\mu\phi}+q_3^\mu q_3^\phi/m_3^2)} {q_3^2-m_3^2}\prod_i {\cal F}_i(m_i, q_i^2)  \ ,\\
{\cal M}_d&=&(i^3)\int \frac{d^4q_2} {(2\pi)^4}[g_{\psi
D^*D}\varepsilon_{\alpha\beta\mu\nu}p_\psi^\alpha\varepsilon_\psi^\beta
(q_3^\nu-q_1^\nu)] [2ig_{D^*D^*V} \varepsilon_V^{*\rho} q_{1\rho}g_{\sigma\tau} +
4if_{D^*D^*V} \varepsilon_V^{*\rho}(2p_{V\tau} g_{\rho\sigma } -p_{V\sigma} g_{\rho \tau})]\nonumber\\
&& \times [2ig_{D^*DP}p_{P\phi} ]\frac {i(-g^{\mu\sigma}+q_1^\mu q_1^\sigma/m_1^2)} {q_1^2-m_1^2}
 \frac {i(-g^{\tau\phi}+q_2^\tau q_2^\phi/m_2^2)} {q_2^2-m_2^2}
\frac {i} {q_3^2-m_3^2}\prod_i {\cal F}_i(m_i, q_i^2)  \ ,\\
{\cal M}_e&=&(i^3)\int \frac{d^4q_2} {(2\pi)^4}[g_{\psi
D^*D^*}\varepsilon_\psi^\alpha(2q_{1\alpha} g_{\beta\mu}+q_{3\beta}
g_{\alpha\mu}-q_{1\mu} g_{\alpha\beta})][-4if_{D^*DV}  \varepsilon_{\rho\sigma\xi\tau}p_V^\rho\varepsilon_V^{*\sigma} q_1^\xi]
 [ig_{D^*DP}p_{P\phi}]\nonumber\\
&&\times \frac {i(-g^{\tau\beta}+q_1^\tau q_1^\beta/m_1^2)} {q_1^2-m_1^2} \frac {i} {q_2^2-m_2^2}
\frac {i(-g^{\mu\phi}+q_3^\mu q_3^\phi/m_3^2)} {q_3^2-m_3^2}\prod_i {\cal F}_i(m_i, q_i^2)  \ ,\\
{\cal M}_f&=&(i^3)\int \frac{d^4q_2} {(2\pi)^4}[g_{\psi
D^*D^*}\varepsilon_\psi^\alpha(2q_{1\alpha} g_{\beta\mu}+q_{3\beta}
g_{\alpha\mu}-p_{1\mu} g_{\alpha\beta})]
\nonumber \\
&&\times  [2ig_{D^*D^*V} \varepsilon_V^{*\rho} q_{1\rho}g_{\sigma\tau} +
4if_{D^*D^*V} \varepsilon_V^{*\rho}(2p_{V\tau} g_{\rho\sigma } -p_{V\sigma} g_{\rho \tau}) ]
 [-i g_{D^*D^*P} \varepsilon_{\phi\lambda\theta\kappa} p_P^\lambda q_2^\theta ]\nonumber \\
& &\times \frac {i(-g^{\beta\sigma}+q_1^\beta q_1^\sigma/m_1^2)}
{q_1^2-m_1^2} \frac {i(-g^{\kappa\tau}+q_2^\kappa q_2^\tau/m_2^2)}
{q_2^2-m_2^2} \frac {i(-g^{\mu\phi}+q_3^\mu q_3^\phi/m_3^2)}
{q_3^2-m_3^2}\prod_i {\cal F}_i(m_i, q_i^2).
\end{eqnarray}
The only parameter of the long-distance part is the cut-off
parameter $\alpha$ which can be determined by two ways. One is to
adopt the same cutoff parameter $\alpha=0.35$ as determined in the
$\psi(3686)$ decays. The other is to determine $\alpha$ by the
branching ratio $BR(\psi(3773)\to J/\psi\eta)=(9\pm 4)\times
10^{-4}$ assuming that this process is also dominated by the IML
with the same cutoff parameter.

\section{Numerical results}
\label{Results}

As discussed in Sec.~\ref{method}, the contributions from the EM and
short-distance part have been fixed by their correlations with the
$\psi(3686)$ decays. There are two parameters to be determined, i.e.
the cut-off parameter $\alpha$ in the charmed meson loop and the
relative phase angle $\theta$ of the meson loop amplitude to the
short-distance part. To evaluate the contributions from the
long-distance part, two ways are used to determine the cut-off
parameter in the charmed meson loop. One way is to use the same
cut-off parameter $\alpha=0.35$ in $\psi(3686)\to VP$ based on the
same topology at the quark level. Since there is no $c\bar{c}$
annihilation short-distance contribution  and the EM transition is
small in $\psi(3770)\to J/\psi\eta$, the long-distance charmed meson
loop is dominant in this process. As a result, the central value
$BR(\psi(3770)\to J/\psi\eta)=(9\pm4)\times 10^{-4}$ gives another
cut-off parameter $\alpha=0.73$. Such a difference indicates that the
vertex form factor which takes care of the non-local effects may
depend on the kinematics strongly as shown in
Ref.~\cite{Guo:2012tj}. Another source of the uncertainties arising from
$\psi(3770)\to J/\psi\eta$ is the SU(3) flavor symmetry breaking.
Note that the coupling for $\psi(3770)$ to $D_s\bar{D}_s$ meson pair
may be different from that for $\psi(3770)D\bar{D}$.
Here we use SU(3) breaking parameter $0.92$  extracted from
universal study of $\psi(3686)$ and $J/\psi$ decays to take account this effect,
i.e. $g_{\psi(3770)D_s\bar{D}_s}/g_{\psi(3770)D\bar{D}}=0.92$ \cite{Wang:2012mf}.
That is also the reason why the cutoff parameter fixed in $\psi(3770)\to J/\psi\eta$ channel
is a little smaller than that in Ref.~\cite{Zhang:2009kr} which does not consider the SU(3) breaking.

In Table~\ref{tab:1} and Table~\ref{tab:2} the branching ratios from
different parts are listed. For the total branching ratios, we
consider two phase angles, i.e. $\theta=0^\circ$ and $180^\circ$,
for each decay modes. It shows that the EM contributions are of
order $10^{-8}\sim 10^{-9}$ which is negligibly small. Meanwhile,
the short-distance and long-distance contributions are both
typically about $10^{-4}\sim 10^{-5}$ for the isospin-conserved
channels, while the long-distance contributions are $10^{-6}\sim
10^{-7}$ for the isospin-violated channels. So it is reasonable to
ignore the EM contributions.

The following points can be learned from the numerical results in
Table~\ref{tab:1} and Table~\ref{tab:2}:
\begin{itemize}
  \item Since the EM interaction plays a minor role in $\psi(3770)\to
  VP$, no significant charge asymmetry can be seen in $KK^*$
  channel.
  \item The branching ratio of $\psi(3770)\to\omega\eta$ from the
  long-distance part
  contribution
  is larger than that of $\omega\eta^\prime$. This is due to the larger $n\bar{n}$
  component in $\eta$. The larger $s\bar{s}$ in $\eta^\prime$ also
  gives the larger long-distance part contribution for $\phi\eta^\prime$ than
  that for $\phi\eta$.
\item The branching ratios of the $\rho\pi$ channel
are found to be larger than those of other channels for both $\alpha$ values.
This is mainly because of the phase space difference. This channel
can be possibly measured by the current BES-III experiment.

\item By summing over all the $VP$ channels,
the first scenario gives the total branching ratio in a range of
$3.83\times 10^{-4}\sim 1.71\times 10^{-3}$. While the second
scenario gives $3.38\times 10^{-2}\sim 5.23\times 10^{-2}$, these
two results set up a range of the inclusive $\psi(3770)\to VP$
branching ratios, which seems to contribute to the $\psi(3770)$
non-$D\bar{D }$ decay significantly.

\end{itemize}

\begin{figure}[tb]
\begin{center}
\includegraphics[scale=0.5]{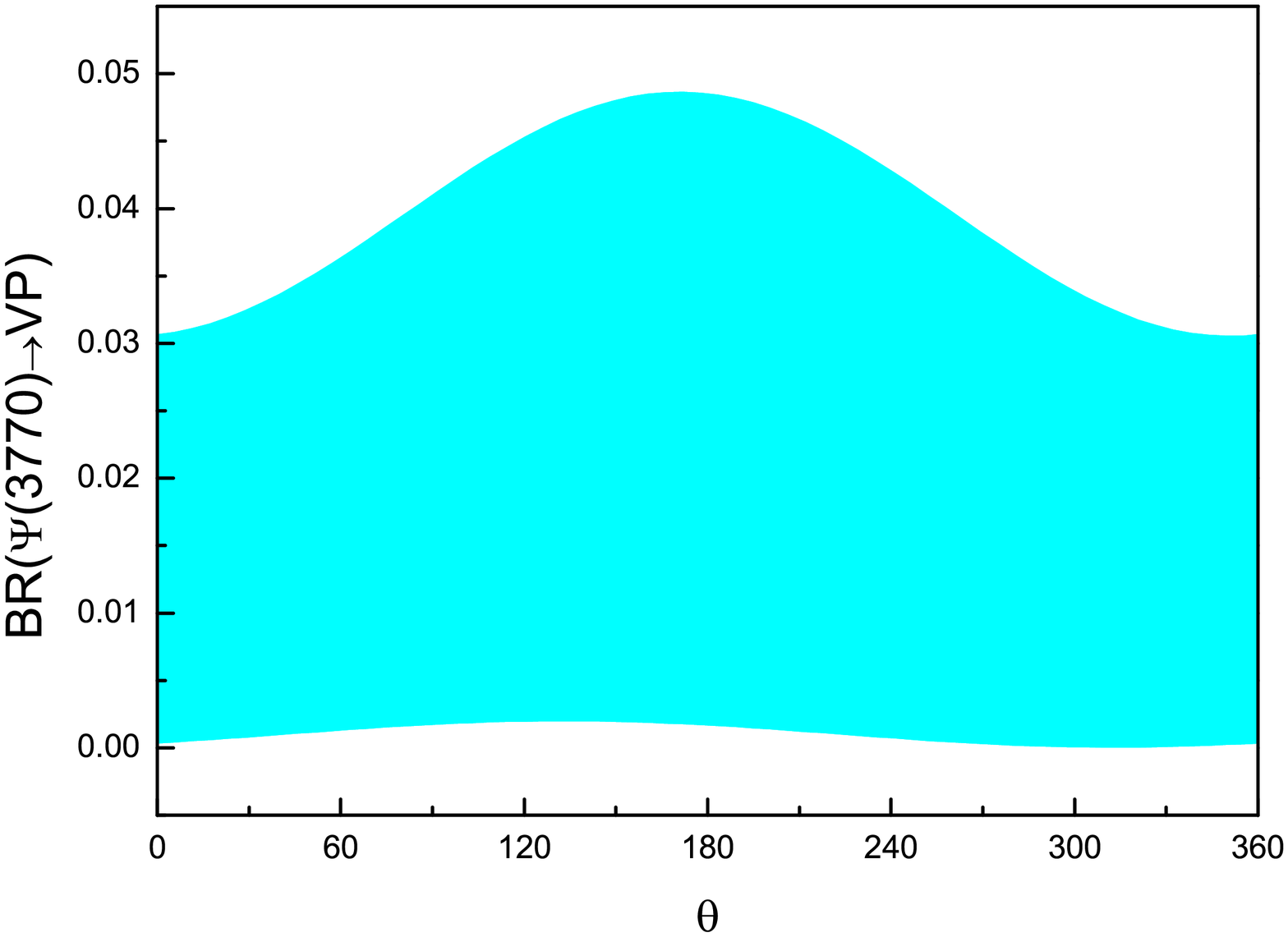}
\caption{The predicted total branching ratios of $\psi(3770)\to VP$
in terms of the relative phase $\theta$ between the short-distance
and long-distance part. The lower limit is obtained with the cut-off
parameter $\alpha=0.35$ determined in $\psi(3686)$ decays, while the
upper limit given by $\alpha=0.73$ is fixed by the branching ratio
$BR(\psi(3770)\to J/\psi\eta)=9\times 10^{-4}$. }\label{phase}
\end{center}
\end{figure}

With the present experimental data, we can not determine the
relative phase between the short-distance and long-distance parts,
although the range of the branching ratios are  comparable with the
recent experimental data~\cite{Besson:2005hm,Ablikim:2007zz}. So we
plot the phase angle dependence of the summed branching ratio in
Fig.~\ref{phase}. The band indicates the range of the summed
branching ratios between $\alpha=0.35$ and 0.73 as an estimate of the
$\psi(3770)$ non-$D\bar{D}$ decays into $VP$, which can be
investigated by the BES-III experiment.

In Ref. \cite{Achasov:2005qb},  the light hadron branching ratios of the $\psi(3770)$
decays from the absorptive part of the intermediated $D\bar{D}$ meson loop are evaluated.
By taking the on-shell approximation and calculating the absorptive part of the $D\bar{D}$ meson loop, we obtain consistent results for $\psi(3770)\to VP$ within the uncertainties of the coupling parameters. It is interesting to note that the closeness of the $\psi(3770)$ to the $D\bar{D}$ mass threshold leads to novel features concerning the locally broken quark-hadron duality. As studied in the literature, for those charmonium states well below or isolated from an open charm threshold, the contributions from intermediate charmed meson loops turns out to be rather small and the sum of those loops generally leads to a cumulative reduction effect~\cite{Achasov:2005qb,Wang:2012wj,Guo:2012tj}. Such an effect may become significant when a threshold is located in the vicinity of the state while other thresholds are relatively far away. For instance, for $\psi(3770)\to D\bar{D}\to VP$
one immediate scenario is that the mass difference between the $D$ and $D^*$ becomes important which apparently violates the heavy quark symmetry between these two states. In this sense, the nontrivial meson loop contributions from the intermediate $D$ meson loops can also be regarded as a manifestation of heavy quark symmetry breaking.  This makes the experimental search for the $\psi(3770)$ non-$D\bar{D}$ decays are extremely important. The importance of the threshold effects can also be seen in the $J/\psi$  and $\psi'$ decays as discussed in Ref.~\cite{Wang:2012mf}.

\begin{table}
\caption{The exclusive and total branching ratios of $\psi(3770)\to
VP$ with the cutoff parameter $\alpha=0.35$ fixed in the
$\psi(3686)$ decays.}
\begin{tabular}{cccccc}
  \hline\hline
  $BR(\psi(3770)\to VP)$ & EM & short-distance  & long-distance & tot(0$^\circ$) & tot(180$^\circ$) \\\hline
  $\rho\eta$ & $2.0\times 10^{-8}$ & 0 & $1.47\times 10^{-6}$ & $1.47\times 10^{-6}$ & $1.47\times 10^{-6}$ \\
  $\rho\eta^\prime$ & $1.17\times 10^{-8}$ & 0 & $6.15\times 10^{-7}$ & $6.15\times 10^{-7}$ & $6.15\times 10^{-7}$ \\
  $\omega\pi^0$ & $3.74\times 10^{-8}$ & 0 & $2.06\times 10^{-6}$ & $2.06\times 10^{-6}$ & $2.06\times 10^{-6}$ \\
  $\phi\pi^0$ & $1.61\times 10^{-10}$ & 0 & 0 & 0 & 0 \\
  $\rho^0\pi^0$ & $7.18\times 10^{-9}$ & $6.86\times 10^{-5}$ & $7.25\times 10^{-5}$ & $7.84\times 10^{-5}$ & $2.04\times10^{-4}$ \\
  $\rho\pi$ & $1.43\times 10^{-8}$ & $2.06\times 10^{-4}$ & $2.17\times 10^{-4}$ & $2.35\times 10^{-4}$ & $6.11\times 10^{-4}$ \\
  $\omega\eta$ & $1.55\times 10^{-9}$ & $4.65\times10^{-5}$ & $5.2\times 10^{-5}$ & $5.42\times 10^{-5}$ & $1.43\times 10^{-4}$ \\
  $\omega\eta^\prime$ & $1.27\times 10^{-9}$ & $1.72\times 10^{-5}$ & $2.21\times 10^{-5}$ & $2.13\times 10^{-5}$ & $5.73\times 10^{-5}$ \\
  $\phi\eta$ & $2.92\times 10^{-9}$ & $1.28\times 10^{-5}$ & $1.4\times10^{-5}$ & $2.63\times 10^{-8}$ & $5.37\times 10^{-5}$ \\
  $\phi\eta^\prime$ & $2.68\times 10^{-9}$ & $2.72\times  10^{-5}$ & $3.30\times 10^{-5}$ & $2.74\times 10^{-7}$ & $1.2\times 10^{-4}$ \\
  $K^{*+}K^-+c.c.$ & $8.59\times 10^{-9}$ & $1.09\times 10^{-4}$ & $9.64\times 10^{-5}$ &$4.35\times 10^{-5}$ & $3.68\times 10^{-4}$ \\
  $K^{*0}\bar{K}^0+c.c.$ & $1.97\times 10^{-8}$ & $1.09\times 10^{-4}$ & $7.82\times 10^{-5}$ & $2.39\times 10^{-5}$ & $3.5\times 10^{-4}$ \\\hline
  total &$1.20\times 10^{-7}$ & $5.27\times 10^{-4}$ & $5.17\times 10^{-4}$ & $3.83\times 10^{-4}$ & $1.71\times 10^{-3}$ \\
  \hline\hline
\end{tabular}
\label{tab:1}
\end{table}

\begin{table}
\caption{The exclusive and total branching ratios of $\psi(3770)\to
VP$ with $\alpha=0.73$ which is fixed by $BR(\psi(3770)\to
J/\psi\eta)=9\times 10^{-4}$.}
\begin{tabular}{cccccc}
  \hline\hline
  $BR(\psi(3770)\to VP)$ & EM & short-distance  & long-distance & tot(0$^\circ$) & tot(180$^\circ$) \\\hline
  $\rho\eta$ & $2.0\times 10^{-8}$ & 0 & $6.30\times 10^{-6}$ & $6.30\times 10^{-6}$ & $6.30\times 10^{-6}$ \\
  $\rho\eta^\prime$ & $1.17\times 10^{-8}$ & 0 & $2.61\times 10^{-6}$ & $2.61\times 10^{-6}$ & $2.61\times 10^{-6}$ \\
  $\omega\pi^0$ & $3.74\times 10^{-8}$ & 0 & $8.89\times 10^{-6}$ & $8.89\times 10^{-6}$ & $8.89\times 10^{-6}$ \\
  $\phi\pi^0$ & $1.61\times 10^{-10}$ & 0 & 0 & 0 & 0 \\
  $\rho^0\pi^0$ & $7.18\times 10^{-9}$ & $6.86\times 10^{-5}$ & $5.60\times 10^{-3}$ & $4.45\times 10^{-3}$ & $6.88\times10^{-3}$ \\
  $\rho\pi$ & $1.43\times 10^{-8}$ & $2.06\times 10^{-4}$ & $1.68\times 10^{-2}$ & $1.34\times 10^{-2}$ & $2.06\times 10^{-2}$ \\
  $\omega\eta$ & $1.55\times 10^{-9}$ & $4.65\times10^{-5}$ & $3.99\times 10^{-3}$ & $3.19\times 10^{-3}$ & $4.87\times 10^{-3}$ \\
  $\omega\eta^\prime$ & $1.27\times 10^{-9}$ & $1.72\times 10^{-5}$ & $1.67\times 10^{-3}$ & $1.35\times 10^{-3}$ & $2.01\times 10^{-3}$ \\
  $\phi\eta$ & $2.92\times 10^{-9}$ & $1.28\times 10^{-5}$ & $9.28\times10^{-4}$ & $7.23\times 10^{-4}$ & $1.16\times 10^{-3}$ \\
  $\phi\eta^\prime$ & $2.68\times 10^{-9}$ & $2.72\times  10^{-5}$ & $2.16\times 10^{-3}$ & $1.70\times 10^{-3}$ & $2.67\times 10^{-3}$ \\
  $K^{*+}K^-+c.c.$ & $8.59\times 10^{-9}$ & $1.09\times 10^{-4}$ & $8.44\times 10^{-3}$ &$6.65\times 10^{-3}$ & $1.05\times 10^{-2}$ \\
  $K^{*0}\bar{K}^0+c.c.$ & $1.97\times 10^{-8}$ & $1.09\times 10^{-4}$ & $8.5\times 10^{-3}$ & $6.69\times 10^{-3}$ & $1.05\times 10^{-2}$ \\\hline
  total &$1.20\times 10^{-7}$ & $5.27\times 10^{-4}$ & $4.25\times 10^{-2}$ & $3.38\times 10^{-2}$ & $5.23\times 10^{-2}$ \\
  \hline\hline
\end{tabular}
\label{tab:2}
\end{table}

\section{Conclusion}
\label{Conclusion}

In this work, we investigate the $\psi(3770)$ non-$D\bar D$ decays
into $VP$ channels which are mainly through the short-distance part
and the long-distance part. The long-distance part is described by
the OZI evading intermediate meson loops according to an effective Lagrangian
method. This calculation enriches and improves our previous
work~\cite{Zhang:2009kr} and confirms the dominance of the IML
contributions in $\psi(3770)\to VP$ via the dispersive part of the
transition amplitude. We make predictions for the branching ratios
of the $VP$ decay channels benefiting from the available measurement
of $\psi(3686)\to VP$ decays and $\psi(3770)\to J/\psi\eta$. These
two experimental constraints provide a range of the $\psi(3770)\to
VP$ decay branching ratios which can be studied by the BES-III
experiment.

It is interesting to see that the IML contributions as a
non-perturbative mechanism indeed account for some deficit for the
$\psi(3770)$ non-$D \bar D$ decays. Such a mechanism is strongly
correlated with the $\psi(3686)$ decays since both states are
located in the vicinity of the $D\bar{D}$ open threshold. Thus, they
will experience large open threshold effects via the IML. This makes
it important to make a coherent study of these processes with the
help of more accurate measurements from experiment. We anticipate
that these results can help us establish the IML as an important
non-perturbative dynamic mechanism in the charmonium decays and
provide insights into the long-standing $\psi(3770)$ non-$D\bar{D}$
decay and ``$\rho-\pi$ puzzle. Further theoretical studies of other
channels such as $\psi(3770)\to VT$, $SP$, $VS$ etc., are also
needed as a prediction or test of the non-perturbative mechanism via
the IML.

\section*{Acknowledgement}
This work is supported, in part, by the National Natural Science
Foundation of China (Grant Nos. 11035006, 11175104, and
11275113, 11121092), the Chinese Academy of Sciences (KJCX3-SYW-N2),
the Ministry of Science and Technology of China (2009CB825200), DFG
and NSFC funds to the Sino-German CRC 110, the Natural Science
Foundation of Shandong Province (Grant Nos. ZR2010AM011 and
ZR2011AM006) and the Scientific Research Starting Foundation of Qufu
Normal University.


\end{document}